\documentclass[article]{elsarticle}
\usepackage{color}

\usepackage{lineno,hyperref}
\usepackage{graphicx}% Include figure files
\usepackage{dcolumn}
\usepackage{amsmath}% Align table columns on decimal point
\usepackage{amssymb}
\usepackage{bm}% bold math
\usepackage{caption}
\usepackage{ulem}
\usepackage{eucal}
\usepackage{mathrsfs}
\usepackage{theorem}
\usepackage{pifont}
\usepackage{sectsty}
\usepackage{amscd}
\usepackage{color}
\usepackage{fancyhdr}
\usepackage{framed}
\usepackage[all]{xy}
\usepackage{enumitem}
\usepackage{multicol}
\modulolinenumbers[5]

\newcommand{\vect}[1]{\mathbf{#1}}

\newcommand{\smatrixII}[3]{\left(\begin{array}{cc}#1&#2\\#2&#3\end{array}\right)}

\newcommand{\dif}[0]{\mathrm{d}}
\newcommand{\dS}{\dif A}

\newcommand{\ab}{\bar{\vect{a}}}
\newcommand{\bb}{\bar{\vect{b}}}
\newcommand{\aac}{\vect{a}}
\newcommand{\bac}{\vect{b}}

\newcommand{\kg}{\kappa_g}
\newcommand{\kgb}{\bar{\kappa}_g}
\newcommand{\Kb}{\bar{K}}
\newcommand{\Lb}{\bar{L}}
\newcommand{\Mb}{\bar{M}}
\newcommand{\Nb}{\bar{N}}
\newcommand{\Gn}[1]{\text{G}_#1}
\newcommand{\MCPn}[1]{\text{MCP}_{#1}}

\newcommand{\figref }[1]{Fig.~\ref{#1}}
\newcommand{\Eqref}[1]{Eq.~\ref{#1}}

\renewcommand{\S}{\mathcal{S}}

\newcommand{\m}{\mathcal{M}}

\newcommand{\id}{\operatorname{Id}}

\def\XXint#1#2#3{{\setbox0=\hbox{$#1{#2#3}{\int}$ }
\vcenter{\hbox{$#2#3$ }}\kern-.6\wd0}}

\newcommand{\beq}{\begin{equation}}
\newcommand{\eeq}{\end{equation}}

\newcommand{\brk}[1]{\left(#1\right)}          % \brk{.}     => (.)
\newcommand{\Brk}[1]{\left[#1\right]}          % \Brk{.}     => [.]
\newcommand{\Abs}[1]{\left| #1 \right|}        % \Abs{.}     => |.|

\newcommand{\GMCP} {Gauss-Mainardi-Codazzi-Peterson }
\newcommand{\MCP} {Mainardi-Codazzi-Peterson }

\newcommand{\stkout}[1]{\ifmmode\text{\sout{\ensuremath{#1}}}\else\sout{#1}\fi}

\journal{Journal of the Mechanics and Physics of Solids}

\begin{document}

\begin{frontmatter}

\title{Hierarchy of Geometrical Frustration in Elastic Ribbons: shape-transitions and energy scaling obtained from a general asymptotic theory}
% Asymptotic theory for geometrically frustrated elastic ribbons

\author[mysecondaryaddress]{Ido Levin\corref{mycorrespondingauthor}\footnote{These authors contributed equally to this work}}
\author[mysecondaryaddress]{Emmanuel Si\'{e}fert$^1$}
\author[mysecondaryaddress]{Eran Sharon}
\author[myfirstaddress]{Cy Maor}
\address[mysecondaryaddress]{Racah Institute of Physics, The Hebrew University of Jerusalem, Israel.}
\address[myfirstaddress]{Einstein Institute of Mathematics, The Hebrew University of Jerusalem, Israel.}

\begin{abstract}
Geometrically frustrated elastic ribbons exhibit, in many cases, significant changes in configuration depending on the relation between their width and thickness.
We show that the existence of such a transition, and the scaling at which it occurs, strongly depend on the system considered.
Using an asymptotic approach, treating the width as a small parameter, 
we find the leading energy terms resulting from the frustration and predict the existence and scaling of the shape transition.
We study in detail 5 different types of frustrated ribbons with a different morphological dependence on ribbon's width: a sharp shape-transition at a critical width, a moderate transition with an intermediate regime, and no transition at all.
We show that the predictions of our approach match experimental results from two different experimental systems: prestressed rubber bilayers and 4D printed thermoplastics, in a wide variety of geometric settings.
\end{abstract}


\end{frontmatter}

%\linenumbers

\section{Introduction}

A ribbon is an elastic body with separation of scale between each of its dimensions: its thickness $t$ is much smaller than its width $w$, which is in turn very small with respect to its length $\ell$ ($t\ll w\ll \ell$).
Such structures may be found in nature, ranging from proteins~\cite{zhang2019shape} to seedpods~\cite{Armon11} or long leaves~\cite{liang2009shape} but also in many man-made systems (saws, belts, mainsprings, etc.).

Ribbons share behavior both with elastic rods (for which $t\sim w$) and with thin-sheets (for which $w\sim \ell$): on one hand, they are slender and floppy like rods, and can be described by one-dimensional models using the width $w$ as another small parameter for dimension reduction; on the other hand, in some regimes their shapes obey strong geometrical constraints, like being (parts of) developable surfaces.
Sadowsky~\cite{sadowsky1930} and then Wunderlich~\cite{wunderlich62} derived such a one-dimensional energy functional for naturally flat and straight ribbons, assuming the ribbon remains stretch-free, and thus forms a  developable surface.
The minimization of the energy functional, to get the equations governing the equilibrium shape under external constraints, was only recently solved by Starostin and van der Heijden~\cite{starostin2007shape} using variational methods and the bicomplex formalism.
This approach was then generalized by Dias and Audoly~\cite{dias2016wunderlich} with a rod-like theory with internal kinematical constraints, tackling also the case of ribbons with geodesic curvature (i.e., ribbons for which the mid-line is not a geodesic curve).
In the mathematical literature, the Sadowsky functional (in fact, a certain correction of it) was rigorously derived from a 2D plate model by Freddi et al.~\cite{FHMP16,FHMP16b}.
All these works assume however that the ribbon remains  stretch-free, which corresponds to the infinitely thin limit.
This makes the analysis impractical for many situations, specifically when relatively large forces are applied to the ribbons or when significant internal stresses are present in the elastic body. 
Very recently, a  nonlinear one-dimensional model was derived in order to account for the stretching of the ribbon's midsurface~\cite{audoly2021one}.
%\Cy{I don't quite understand the last sentence: we expect the Sadowsky functional to be more or less fine even for incompatible bodies if the width is large enough compared to the thickness (e.g., as in Efi's paper), whereas even when there are very small forces, but the ribbon is narrow enough, then Sadowsky cannot be expected to be useful.}

When stretched and twisted, for example, ribbons exhibit very rich morphologies: smooth and wrinkled
helicoids, cylindrical or faceted patterns may be observed depending on the geometry and the applied stretch and twist~\cite{green36,dinh2016,chopin2013helicoids,chopin2016roadmap}.
%\sout{Some of these configurations are not stretch-free, hence cannot be treated using the above models}.
Shape transitions and rich morphologies are also observed in pre-stressed, frustrated ribbons that are \emph{ free of external loading}~\cite{selinger2004shape,bruinsma2005,Armon11,santangelo2009buckling,sawa2011shape,armon14,huang2018}.
This frustration has been described geometrically in the framework of incompatible elasticity, where frustration stems from the incompatibility of the reference geometry.
First theoretical principles were developed in three-dimensional elastic bodies to tackle the problem of growth~\cite{KondoBook,amar2005growth}. 
For slender sheets, however, a dimensional reduction may be performed, using the thickness $t$ as a small parameter.
Then, the body is described as a $t/2$-tubular neighborhood of a mid-surface $\S$ (i.e., the set of points whose distance is at most $t/2$ from the mid-surface)
~\cite{Dervaux08,dervaux2009morphogenesis,efrati2009elastic}.
The dimensionless Hookean elastic energy functional may be separated into two terms measuring stretching and bending energies ($S$ and $B$, respectively)\cite{Armon11,grossman16}:
\begin{equation}
    W_{2D}\equiv S + B =\dfrac{1}{A}\int(\aac-\ab)^2\dS+\dfrac{t^2}{A}\int(\bac-\bb)^2\dS
    \label{eq:energy}
\end{equation}
where $t$ is the thickness, $(\ab,~\bb)$ are the reference metric and curvature tensors, $(\aac,~\bac)$ are the actual metric and curvature tensors adopted by the structure, and $A$ is the surface area in the reference frame (i.e., $\dS\equiv\sqrt{\Abs{\ab}} \dif x_1 \dif x_2$ is the reference area measure). 
The first term measures the stretching energy in the body, whereas the second term represents the bending energy.
Here, for the sake of simplicity, we non-dimensionalized the elastic energy by dividing it by the quantity $8AtY/(1-\nu^2)$, where $(Y,\nu)$ are the Young's modulus and Poisson ratio of the material.
The  terms $(\aac-\ab)^2$ and $(\bac-\bb)^2$ are calculated using the elastic tensor of $\ab$. The exact form, which depends on the Poisson ratio, is immaterial for this work and appears in \cite{grossman16} Eq.\ (1).
The tensor fields $(\aac,~\bac)$, being the actual metric and curvature tensors of the body in Euclidean space, are not independent; they are related by the \GMCP equations \cite{Ciarlet05,do1992riemannian}.
On the other hand, the reference fields $(\ab,~\bb)$ do not necessarily come from fields of an immersed surface, and thus, might not satisfy these equations. 
When they are not, no surface can simultaneously satisfy both the reference metric and curvature fields. 
In these cases, the energy may not vanish, and the structure is pre-stressed (i.e., frustrated).\footnote{The fact that for incompatible $\ab$ and $\bb$ the infima of energies of the type $W_{2D}$ do not vanish was rigorously proved recently by Alpern et al.\ \cite{AKM20}.}
Many ribbons were shown to be frustrated according to Gauss' equation, which states that for a 2D surface, $\det \bac / \det \aac$ is equal to  the Gauss curvature $K$ of $\aac$ which is determined by the Brioschi formula (see \ref{ap:metric}).
Hence, Gauss frustrated ribbons are structures whose reference geometry satisfies $\det \bb / \det \ab \ne \Kb$.

Among these are chiral seedpods~\cite{Armon11} which have a negative reference curvature but a flat metric, but also macromolecules~\cite{zhang2019shape} (Fig.\ 1a). 
Naturally flat ribbons with non-Euclidean metrics \cite{santangelo2009buckling,grossman16,efrati2016non} buckle out of plane when sufficiently thin.
In all these examples, two asymptotic regimes may be identified: in a stretching-dominated limit, i.e., for thin and wide ribbons, the ribbon adopts a shape which is an embedding of the reference metric, at bending costs: isometries are thus favored.
In the bending-dominated regime, i.e., for thick and narrow ribbons, the ribbon obeys the reference curvature at stretching cost.
A sharp transition, known as Euler buckling, separates both regimes \cite{efrati2009buckling}.
It may be estimated by equating both bending and stretching energy terms.
The typical strains resulting from a mismatch in Gaussian curvature, $K$, scale like $w^2 K$ and hence the stretching energy scales like $w^4 K^2$~\cite{Armon11,armon14}.
Then, from the typical bending energy scaling $t^2 K$
we get the scaling law for the transition:
\begin{equation}\label{eq:transition_Gauss}
    w^4\sim t^2/K
\end{equation}
An example for such transition is treated in detail in Section \ref{sec:nep}.

\begin{figure}[h!]
    \centering
    \includegraphics[width=\linewidth]{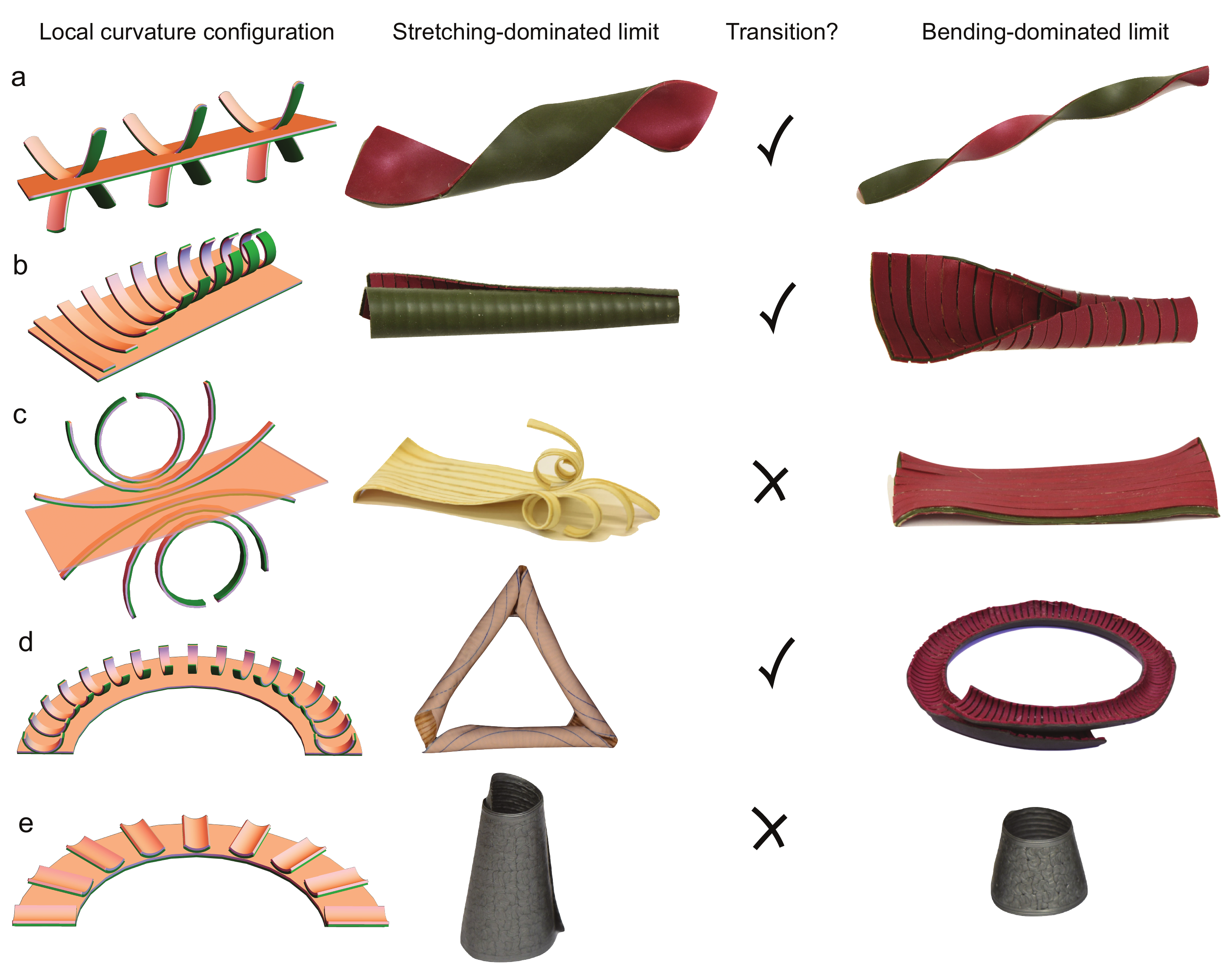}
    \caption{{\bf Examples of geometrically frustrated ribbons}.
    (a) Straight ribbon with a flat metric and constant negative Gaussian reference curvature, as presented in \cite{Armon11}. The principal directions of curvature form an angle of $\pm45^\circ$ with respect to the longitudinal axis. The ribbon sharply transitions between two helical con?gurations: one cut from a cylinder in the stretching-dominated regime, and a helicoid in the bending-dominated regime (See Section~\ref{sec:nep}).
    (b) Straight ribbon with linearly varying transverse reference curvature along the long direction. 
    In the stretching dominated regime, the ribbon rolls into a conical shape whereas it morphs onto a trumpet-like shape in the bending-dominated regime (Section~\ref{sec:trumpet}).
    (c) Straight ribbon with linearly varying reference curvature along the width.
    Independently of the geometry of the ribbon and the amplitude of the spontaneous curvatures, the ribbon remains flat, except in a small region near both ends of the ribbon.
    The ribbon in the stretching-dominated regime has been cut in fine strips at one end to show its reference curvature (Section~\ref{sec:invtrumpet}).
    (d) Curved ribbon with constant radial reference curvature $\kappa$.
    The ribbons transition from polygonal flat structures to toroidal shapes~\cite{PRX} (Section~\ref{sec:radial}).
    (e) Curved ribbon with constant azimuthal reference curvature $\kappa$.
    The ribbons coils into conical shapes, irrespective of the geometrical parameters (Section~\ref{sec:azimuthal}).
    The analysis of these systems is summarized in Table~\ref{tab}.
    }
    \label{fig1}
\end{figure}

It was recently shown that geometric frustration could also arise from incompatibility in the \MCP equations \cite{PRX}.
For a ribbon with a flat reference metric, a change of orientation or magnitude of a unidirectional reference curvature does not contradict Gauss equation, since the curvature tensor $\bb$ remains locally uniaxial, and hence its determinant is zero. % of the curvature tensor is zero. 
It is however in general incompatible with the strong geometrical constraints imposed on developable surfaces.
In such cases, the induced frustration comes from the two other compatibility equations, i.e., the \MCP (MCP) equations. 
Four examples of MCP-frustrated ribbons are shown in Fig.\ 1.
In Fig.\ 1b and c, a straight ribbon has a longitudinal (respectively transverse) unidirectional curvature of varying amplitude in the orthogonal direction.
In Fig.\ 1d and e, ribbon with reference geodesic curvature $\kgb$ has a constant radial respectively azimuthal reference curvature.
Despite their similarities, these examples surprisingly exhibit very different responses: examples {\it b} and {\it d} show shape transitions and rich morphologies, whereas examples {\it
c} and {\it e} lead to a unique shape, without any transition.
As we shall show later, examples {\it b} and {\it d} do not obey the same scaling law for the transition, and both scaling laws are different from the one of  Gaussian frustrated ribbons.

In this article, we present and derive a systematic asymptotic theory to rationalize the shapes and transitions observed. 
Expanding around the midline, we solve order by order the compatibility equations and identify the leading order in $w$ of the frustration.
This immediately provides bounds on the scaling of the transition, as well as information on the expected configurations.
These include the Gauss-incompatibility transition \eqref{eq:transition_Gauss}, as well as MCP-incompatibilities and higher order Gauss-incompatibilities.
We first introduce the general asymptotic theory, explaining the systematic procedure to identify the leading order of the competition between bending and stretching energies. 
We then apply this procedure to the five model examples presented in Fig.\ 1, and compare the scaling of the transition with experiments. 
The theory shows remarkable agreement with the different configurations studied, and we further refine our analysis by finding developable shapes with lower bending energy in the stretching-dominated limit, leading to tighter bounds on the transition scaling.

\section{Asymptotic theory}
\label{sec:asymptotic}

\begin{figure}[h!]
    \centering
    \includegraphics[width=0.75\linewidth]{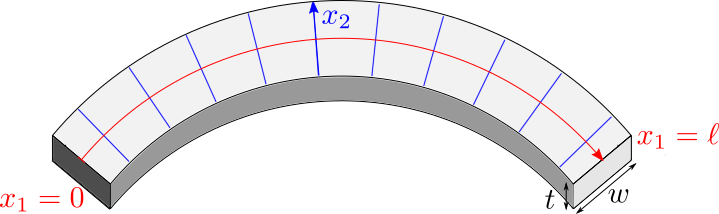}
    \caption{{\bf Construction of semigeodesic coordinates}.
    An illustration of a curved ribbon of length $\ell$, width $w$, and thickness $t$.
    The $x_1$ coordinate (red) is an arclength coordinate along the midline.
    The $x_2$ coordinate (blue) is constructed by arclength geodesics that are perpendicular to the midline.
    }
    \label{fig_coord}
\end{figure}

Let us consider an elastic ribbon $\S$, that is a $\dfrac{w}{2}$-tubular neighborhood of some midline $\m\subset \S$, with a reference geometry encoded by $\ab$ and $\bb$.
One may parametrize the ribbon by choosing the $x_1$-coordinate to be an arclength coordinate along the midline $\m$, and $x_2$ to be the arclength coordinate of $S$-geodesics emanating from $\m$ and perpendicular to it (\figref{fig_coord}).
With this parametrization, the metric tensor reads:
\[
\ab = \smatrixII{\ab_{11}}{0}{1}
\]
From Gauss' \emph{Theorema Egregium}, we know that every metric tensor has an associated intrinsic Gaussian curvature.
For this parametrization of the reference metric, the associated reference Gaussian curvature reads~\cite{struik1961}:
\beq
    \Kb = \dfrac{(\partial_2 \ab_{11})^2 - 2 \ab_{11} \partial_2^2 \ab_{11}}{4 \ab_{11}^2}
    \label{eq:Kbar}
\eeq
where $\partial_i$ signifies a derivative with respect to $x_i$.
We expand both reference metric and Gaussian curvature as power series of $x_2$:
\begin{align*}
\ab(x_1,x_2) &= \ab_0(x_1) + \ab_1(x_1) x_2 + \ab_2(x_1) x_2^2 + O(x_2^3)\\
\Kb(x_1,x_2) &= \Kb_0(x_1) + \Kb_1(x_1) x_2 + \Kb_2(x_1) x_2^2 + O(x_2^3).
\end{align*}
Note that $\ab_0=\id$ since the $x_1$ is an arclength coordinate. We now identify the geometric meaning of $\ab_1$. Recall that the geodesic curvature $\kgb$ of a line in a surface can be defined as the derivative along the surface of its normal (i.e., its deviation from being a geodesic). From this, a simple calculation shows that $\ab_{11} = 1 - 2\kgb x_2 + O(x_2^2)$.
Now, explicitly calculating  \Eqref{eq:Kbar} and comparing both expansions of $\Kb$ order by order, we can expand the reference metric using $\kgb$ and the expansion of $\Kb$:
Then, we identify the corresponding coefficients and get $\ab_{11}$ as a function of both the geodesic curvature $ \kgb(x_1)$ of the midline $\m$ on $\S$ and $\Kb$:
\[
\begin{split}
\ab_{11}    &= 1 - 2\kgb x_2 + (\kgb^2 - \Kb_0) x_2^2 + \dfrac{1}{3}(4\kgb\Kb_0 - \Kb_1) x_2^3 \\
            &\quad + \dfrac{1}{6} \brk{2\Kb_0^2 -\Kb_2 + 3\Kb_1 \kgb - 2\Kb_0 \kgb^2} x_2^4 + O(x_2^5).
\end{split}
\]
\noindent
We similarly expand the reference curvature
\[
\bb(x_1,x_2) = \bb_0(x_1) + \bb_1(x_1) x_2 + \bb_2(x_1) x_2^2 + O(x_2^3),
\]
and other quantities that appear below.
We denote $\bb = \smatrixII{\Lb}{\Mb}{\Nb}$.

Frustration arises whenever $\ab$ and $\bb$ do not satisfy either the Gauss (G) or the Mainardi--Codazzi--Peterson (MCP) equations, which in this case of semi-geodesic coordinates takes a simpler form:
\begin{equation}
    \begin{split}
        \text{G}&:\ab_{11}\Kb  = \Lb\Nb - \Mb^2
        %\Gn{\,}&:(\partial_2 \ab_{11})^2 - 2 \ab_{11} \partial_2^2 \ab_{11} = 4 \ab_{11}(\Lb\Nb - \Mb^2)
        \\
        \text{MCP}&:\left\{
        \begin{split}
           2 \ab_{11}(\partial_2 \Lb-\partial_1 \Mb) &=  \partial_2\ab_{11}\Lb - \partial_1\ab_{11} \Mb + \ab_{11} \partial_2\ab_{11} \Nb
           \\
           2 \ab_{11}(\partial_1 \Nb-\partial_2 \Mb) &= \partial_2\ab_{11} \Mb
        \end{split}
        \right.
    \end{split}
    \label{eq:full_GMCP}
\end{equation}
If \eqref{eq:full_GMCP} is not satisfied, then there are no metric and curvature tensors $\aac,\bac$ of a surface in $\mathbb{R}^3$ satisfying $\aac = \ab$ and $\bac = \bb$ simultaneously.
Thus, in this case, the energy \eqref{eq:energy} cannot vanish.

We now expand the energy \eqref{eq:energy} order-by-order in $x_2$ and integrate in the $x_2$ direction, resulting in an effective 1D energy of the type 
\begin{equation}\label{eq:energy_1D}
    \begin{split}
      W_{1D}    &= \int_\ell \brk{\alpha_*|\aac_{0} - \ab_{0}|^2 + \alpha_g w^2 |\aac_{1} - \ab_{1}|^2 + \alpha_0 w^4 |\aac_{2} - \ab_{2}|^2 + \ldots}\,dx_1 \\
                &\quad + t^2 \int_\ell \brk{\beta_0|\bac_{0}- \bb_{0}|^2 + \beta_1 w^2 |\bac_{1} - \bb_{1}|^2 + \ldots}\,dx_1 \\
                &\equiv \int_\ell \brk{S_*(x_1) + S_g(x_1) + S_0(x_1) + \ldots}\,dx_1 \\
                &\quad + \int_\ell \brk{B_0(x_1) + B_1(x_1) + B_2(x_1) + \ldots}\,dx_1,
    \end{split}
\end{equation}
where the $S_*\equiv \alpha_*|\aac_{0} - \ab_{0}|^2$ term results from stretching the midline, the $S_g \equiv \alpha_g w^2 |\aac_{1} - \ab_{1}|^2$ term from a deviation in the geodesic curvature,
the $S_k\equiv w^{4+2k}\alpha_k  |\aac_{k+2} - \ab_{k+2}|^2$ terms (for $k\ge0$) result from further expanding the stretching term $S$, and the $B_j\equiv t^2 w^{2j} \beta_j  |\bac_{j} - \bb_{j}|^2$ terms result from expanding the bending term $B$ in \eqref{eq:energy} (the reason for the numbering of the terms will be clear from the analysis below).
In this expansion, we omit the mixed terms, such as $(\aac_{0} - \ab_{0})\cdot(\aac_{2} - \ab_{2})$, since we are only interested in the \emph{leading} non vanishing terms in the stretching and bending energies;
the omitted terms involve previous terms in the expansion and hence are never leading order terms. 
In addition, once we express both metrics using the corresponding Gaussian curvatures, assuming all the lower stretching terms vanish, we have $S_k \sim w^{4+2k}|K_k-\Kb_k|^2$, that is, $S_k$ is the energetic contribution of the $k$th order deviation of the Gaussian curvature.

Assuming $\ab$ and $\bb$ do not satisfy \eqref{eq:full_GMCP}, we aim to minimize the energy order-by-order, and find which terms are competing and cannot vanish simultaneously.
Since $t\ll w$, the first two terms in the stretching energy, $S_*$ and $S_g$, are more expensive than any of the bending terms; minimizing them thus results in $\aac_0 = \ab_0$ and $\aac_1 = \ab_1$, that is,
\[
\aac = \smatrixII{1-2\kgb(x_1) x_2}{0}{1} + O(x_2^2).
\]
In particular, as the linear term in $\aac_{11}$ is (minus $2$ times) the geodesic curvature of the configuration, we obtain that the ribbon maintains the reference geodesic curvature of the midline (i.e., $\kg = \kgb$)~\cite{grossman16}.
Higher terms in $\aac$ may result in stretching energy comparable to the lowest bending energy terms.
Hence, they are potentially competing with $\bac$ through the compatibility equations.
In order to identify and characterize the frustration we expand these equations order by order:
\begin{align}\label{eq:Gn}
        \Gn{0} &: K_0=L_0 N_0-M_0^2 \nonumber\\
        \Gn{1} &: K_1-2 \kgb K_0 =L_1 N_0+L_0 N_1-2 M_0 M_1 \\
        \Gn{2} &: K_2 -2\kgb K_1  + \kgb^2 K_0  - K_0^2=L_2 N_0 + L_1 N_1 + L_0 N_2 - M_1^2-2 M_0 M_2 \nonumber
\end{align}
for Gauss equation, and 
\begin{align}\label{eq:MCPn}
    \MCPn{0} &:
    \begin{array}{l}
         L_1=M_0'-\kgb (L_0+N_0) \\
         M_1=N_0' +\kgb M_0  
    \end{array}
 \\
    \MCPn{1} &:
    \begin{array}{l}
        2 L_2= M_1' +\kgb' M_0 + \kgb^2 (N_0-L_0)-\kgb (L_1+N_1) -K_0 (L_0+N_0) \\
        2 M_2= N_1' +\kgb^2 M_0 +\kgb M_1 + K_0 M_0
    \end{array}\nonumber
\end{align}
for the \MCP equations (all the derivatives are taken with respect to $x_1$).
The equation $\MCPn{0}$ arises from the fact that $\aac_0 = \ab_0$ and $\aac_1 = \ab_1$. 
Similarly, if we have that $\aac$ is of the semi-geodesic form $\smatrixII{a_{11}}{0}{1}$ up to order $x_2^k$, then the Gauss equation and the MCP equations take the form  \eqref{eq:full_GMCP} up to order $k-2$ and $k-1$ in $x_2$, respectively;
this is because the $i$th order of the Gauss equation involves  derivatives of $\aac$ up to order $i+2$, and the $j$th order of the MCP equations involves derivatives up to order $j+1$.
In other words, the $\MCPn{j}$ and $\Gn{i}$ equations take the form of expansions of \eqref{eq:full_GMCP} for orders $j\le k-1$ and $i\le k-2$, respectively.
Note that the semi-geodesic form of $\aac$ is not an additional assumption but rather a result of the order-by-order minimization, as explained in \ref{ap:metric}.

An immediate result of these expansions is that the zeroth-order Gauss equation $\Gn{0}$ is the leading constraint since it only involves terms in $S_0$ and $B_0$. 
This is why, the zeroth order Gauss frustration, which is the one commonly studied in the literature, is dominant over MCP frustration for ribbons.
Next, $\Gn{1}$ and $\MCPn{0}$ contain both zeroth and first-order terms.
When possible, the first-order terms are selected to satisfy $\Gn{1}$ and $\MCPn{0}$ under the constraint on the zeroth-order terms set by $\Gn{0}$.
The same logic follows to the higher order compatibility conditions, where the terms involved in $\Gn{j}$ and $\MCPn{j-1}$, and not constrained by previous orders, contribute to $S_j$ and $B_j$.

We now present an algorithm to determine the leading energy terms and the existence and scaling of a shape transition from the asymptotic expansion.
Several specific examples for the use of this algorithm appear in \ref{sec:examples}.
\begin{itemize}
    \item \textbf{Finding the leading bending frustration:}
        When the ribbon is sufficiently wide (i.e., in a stretching-dominated regime), we expect the configuration to be an isometric embedding of the reference metric (i.e., $\aac=\ab$) \cite{kupferman2014}.
        Furthermore, in order to minimize the bending energy, $\bac$ will approximate $\bb$, as long as it does not violates the GMPC equations (for $\aac=\ab$).
        Therefore, in order to approximate the bending energy in this limit we successively check the self consistency of $\bac_j = \bb_j$ under the compatibility conditions and the isometric constraint, $\aac=\ab$.
        The first inconsistent order $j$ will result in a leading energy term $ B_j \sim t^2 w^{2j} \Abs{\bac_j-\bb_j}^2$.
    \item \textbf{Finding the leading stretching frustration:}
         The inconsistency between $\aac=\ab$ and $\bac_j=\bb_j$ tells us the leading stretching term as well. The first term $K_k$ which is inconsistent with $\bac_j = \bb_j$, that is, the leading order in $K$ which, if relaxed, is consistent with $\bac_j = \bb_j$.
        Such a relaxation leads to stretching energy $S_k \sim w^{4+2k}\Abs{K_k-\Kb_k}^2$.
    \item \textbf{Determining the existence and scaling of the shape transition:}
         Once the leading orders are determined, we compare their scaling to find the shape transition. 
         If $k\ge j$, the frustration is between $S_k$ and $B_j$ and the scaling of the transition is expected to be $t \propto w^{2+k-j-\gamma}$, where $0\le \gamma < 1$.
        The correction $\gamma$ results from the possibility to perturb the terms $\bac_{i}$ for $i<j$ around $\bb_i$, and thus lowering the energy while still satisfying the compatibility equations.
        This is what happens in Fig.\ 1a, 1b and 1d, as we show below.
        %, which provide perturbative approach which provides only bounds between the terms of the expansion.
        
        However, since asymptotically $S_j > B_{j+1}$ (as $t\ll w$), if $k<j$ a phase transition might not occur, in which case the wide limit is the only viable configuration.
        This is exactly what happens in Fig.\ 1c and 1e, as we show below.
\end{itemize}
Therefore, our approach provides also a selection rule for the existence of a shape transition in frustrated ribbons.\footnote{In principle there might be phase transitions also when $k<j$, due to competition between higher order stretching and bending, but even then the transition will be more subtle as it will involve higher orders in the configuration.}

\section{Examples}
\label{sec:examples}
Using the procedure described in the previous section, we now tackle the five examples presented in Fig.~\ref{fig1}. Details on the fabrication of the different samples may be found in \ref{ap:fab}. We shall show that each example exhibits different leading orders in the incompatibility between bending and stretching terms, leading to different scalings of the shape transition. 
Furthermore, while the stretching dominated regime obeys $\aac = \ab$ (i.e., isometries are favored in the thin and wide limit), the bending energy may not vanish even in the bending dominated regime, due to \MCP incompatibility.

\subsection{Gauss zeroth-order incompatibility: frustration between $S_0$ and $B_0$}
\label{sec:nep}
This particular case was already extensively studied~\cite{selinger2004shape,efrati2016non,chen2014}, in the context of seed pods \cite{Armon11, armon14}, macromolecules~\cite{zhang2019shape} or anomalously soft ribbons~\cite{guest11,levin16} and rationalized using a very similar approach in \cite{grossman16}, that we shortly recapitulate.

For concreteness we consider the specific case that appears in \figref{fig1}a: a straight ribbon parametrized by $0 \le x_1\le \ell$ and $-w/2\le x_2\le w/2$, having a flat reference metric and  a constant of-diagonal reference curvature, $\kappa$: 
\[
\ab = \smatrixII{1}{0}{1} 
\qquad
\bb = \smatrixII {0}{\kappa}{0}
\]
In this case $\Kb\equiv 0$, but $\Lb_0 \Nb_0-\Mb_0^2=-\kappa^2\ne0$ hence we have a zeroth-order Gauss incompatibility.
Starting with $\Gn{0}$ in the wide-limit, we find $L_0 N_0-M_0^2=0$.
Minimizing $B_0$ results in all components of $\bac_0-\bb_0$ proportional to $\kappa$, corresponding to a helical shape.
Hence, we find $B_0\sim t^2 \kappa ^2$.

Since the frustration involves $K_0$, the leading stretching term is $S_0$.
Obeying $\bb_0$ results in $K_0=-\kappa^2$, hence $S_0\sim w^4\kappa^4$.
Finally, we compare both terms to find a shape transition:
\[
t\sim w^2 \kappa
\]
Whenever $\Gn{0}$ is satisfied however, frustration will result from higher order terms and transitions with atypical scaling laws may appear, as we will see in the next examples.

\subsection{Linearly increasing transverse curvature: frustration between $S_2$ and $B_0$}
\label{sec:trumpet}

Consider a straight ribbon parametrized by $\ell_1-\ell \le x_1\le \ell_1$ and $-w/2\le x_2\le w/2$, having a flat reference metric and  a linearly varying transverse curvature, with a typical lengthscale, $\lambda$ (\figref{fig1}b): 
\[
\ab = \smatrixII{1}{0}{1} 
\qquad
\bb = \smatrixII {0}{0}{\lambda^{-2} \, x_1}
\]

In this case $\Kb\equiv 0$, and $\ab$ and $\bb$ satisfy Gauss equation (everywhere).
However, $\ab$ and $\bb$ are incompatible since the \MCP equations are not satisfied.
Following our scheme, we assume $K=\Kb$ and check order by order $\bac_0 = \bb_0$ for consistency.
Starting with $\Gn{1}$ and $\MCPn{0}$ we find
\[
\begin{split}
    \MCPn{0}&:\quad \left\{
    \begin{split}
         L_1 &= 0\\
 M_1 &= \lambda^{-2} 
    \end{split}
    \right.
    \\
\Gn{1}&:\quad L_1 = 0
\end{split}
\]
which is consistent and provides constraints on the next order.
However, considering $\Gn{2}$ and $\MCPn{1}$ yields
\begin{equation}
\begin{split}
    \MCPn{1}&:\quad \left\{
    \begin{split}
       2 L_2 &= 0\\
       2 M_2 &= N_1'
    \end{split}
    \right.
    \\
\Gn{2}&:\quad L_2 = \lambda ^{-2} x_1^{-1}
\end{split}
\label{eq:S2B0}
\end{equation}
which is inconsistent.
The assumption leading to the inconsistency is $\bac_0 = \bb_0$, and thus the dominant bending energy term is $B_0$.
Furthermore, the contradiction obtained involves $\Gn{2}$ (which can be written as $K_2 = \lambda ^{-2} x_1 L_2 -\lambda ^{-4}$). 
Thus, in order to satisfy $\bac_0 = \bb_0$, the constraint $K_2 = \Kb_2=0$ should be relaxed, hence the leading stretching energy term is $S_2$.
We finally obtain a competition between $ B_0$ and $S_2$ energy terms, that is, between terms of order $t^2$ and $w^8$, respectively, suggesting a phase transition at $t\propto w^4$.
Elaborately, satisfying $K_0 = \Kb_0=0, K_1 = \Kb_1=0$ and $b_0 = \bb_0$ simultaneously implies $K_2 = -\lambda^{-4}$ (from equation \eqref{eq:S2B0}). The leading stretching energy term thus scales as $S_2 \sim w^8 \lambda^{-8}$.
On the other hand, satisfying the vanishing reference Gaussian curvature up to the second-order imposes $\bac_0 \ne \bb_0$, leading to a bending energy density $\Abs{\bac_0 - \bb_0}^2\sim \lambda^{-4}x_1^2$.
Hence, the bending energy typically scales as $B_0 \sim t^2 \lambda^{-4} \ell^2$.
Balancing both stretching and bending leading order energies, we get the following scaling for the transition:
\begin{equation}
  t \sim \ell^{-1} \lambda ^{-2} w^4  
  \label{eq:w4}
\end{equation}

As discussed above, this scaling law is actually an upper bound, hence $t \propto w^{4-\gamma}$ where $0\le\gamma<1$.

Searching for a tighter bound, we aim to perturb $\bac_0$ around $\bb_0$, so that we do not violate the compatibility equation $G_2$. We observe that the following metric and curvature tensors
\beq\label{eq:ab_perturb}
\aac = \smatrixII{1}{0}{1}
,\qquad
\bac = \lambda^{-2}\left(x_1 + \dfrac{1+\epsilon}{\delta} x_2\right) \smatrixII{\delta^2}{\delta (1+\epsilon)}{(1+\epsilon)^2}
\eeq
 satisfy the compatibility equations for any fixed $\delta$ and $\epsilon$.
We assume that $\Abs{\delta},\Abs{\epsilon}\ll 1$ and search for the values that minimize the bending content (since the stretching content vanishes).
The bending content per unit volume is of order
\[
\dfrac{\delta ^2 \ell^2 (2 \epsilon +1)}{36 \lambda ^4}+\dfrac{w^2 (1-6 \epsilon )}{48 \delta ^2 \lambda ^4}
\]
which is minimized for $\delta^2 \sim w/\ell$ and $\epsilon \sim w/\ell$, resulting in a bending energy density of order $B_\delta \sim t^2 w \lambda^{-4} \ell$.
The transition scaling described in \eqref{eq:w4} is thus slightly modified and reads 
\begin{equation}
t \sim \ell^{-1/2}\lambda^{-2}  w^{7/2}
\end{equation}
\begin{figure}[h!]
    \centering
    \includegraphics[width=\linewidth]{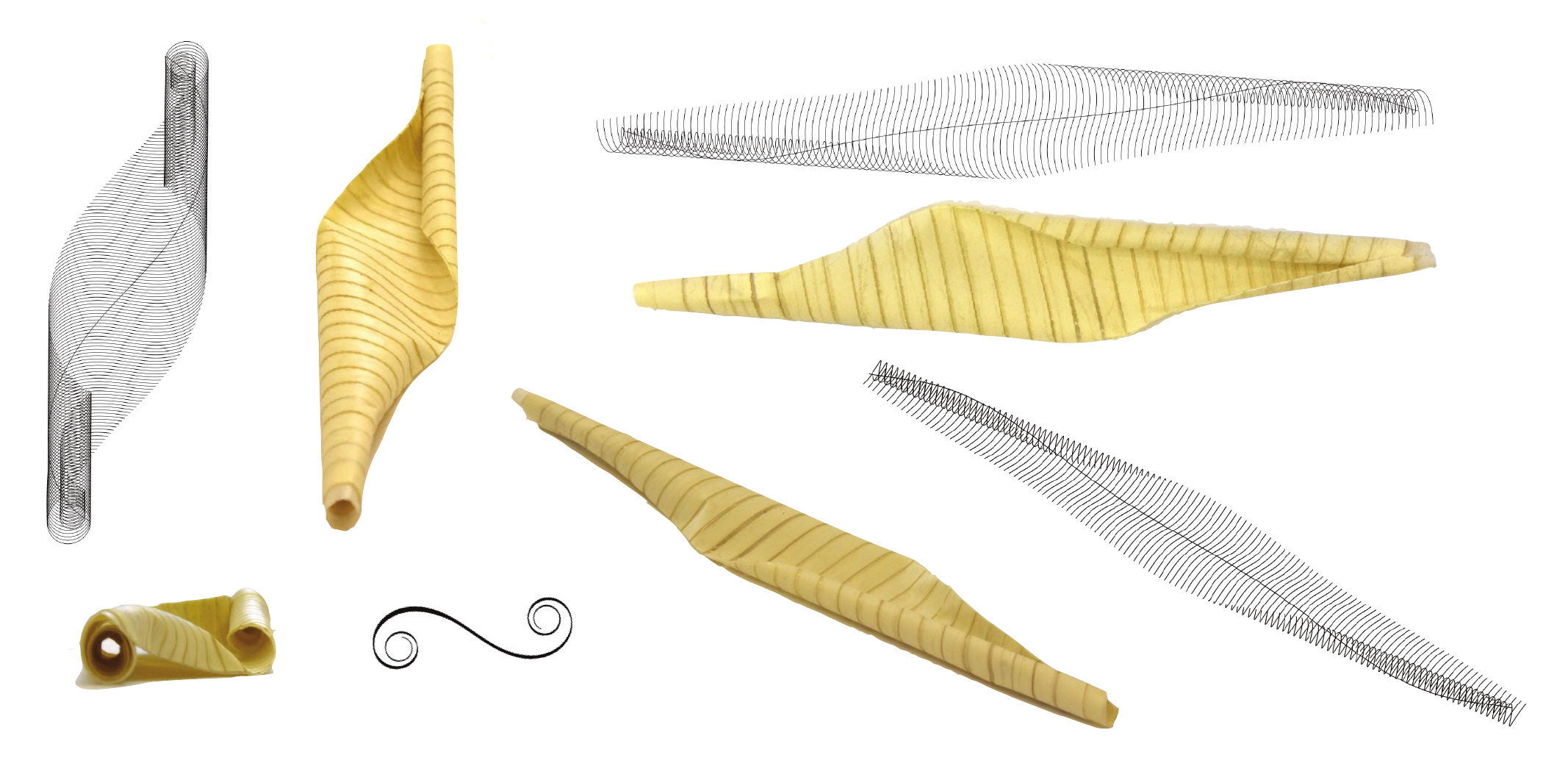}
    \caption{{\bf The shape of a ribbon with linearly varying reference curvature in the stretching-dominated regime.} The curvature of the ribbon varies along the long dimension, switching its sign. Different viewpoints of a single ribbon, comparing the theoretical shape integrated from the proposed metric and curvature tensors (Equation \ref{eq:ab_perturb}) and an experimental sample. They show very good agreement, except near the edges, where a typical boundary layer with Gaussian curvature is present. Experimental parameters: $\ell=190$ mm, $t=0.05$ mm, $w=30$ mm and $\lambda=14$ mm)
    }
    \label{figtrumpetiso}
\end{figure}
 \begin{figure}[h!]
    \centering
    \includegraphics[width=0.9\linewidth]{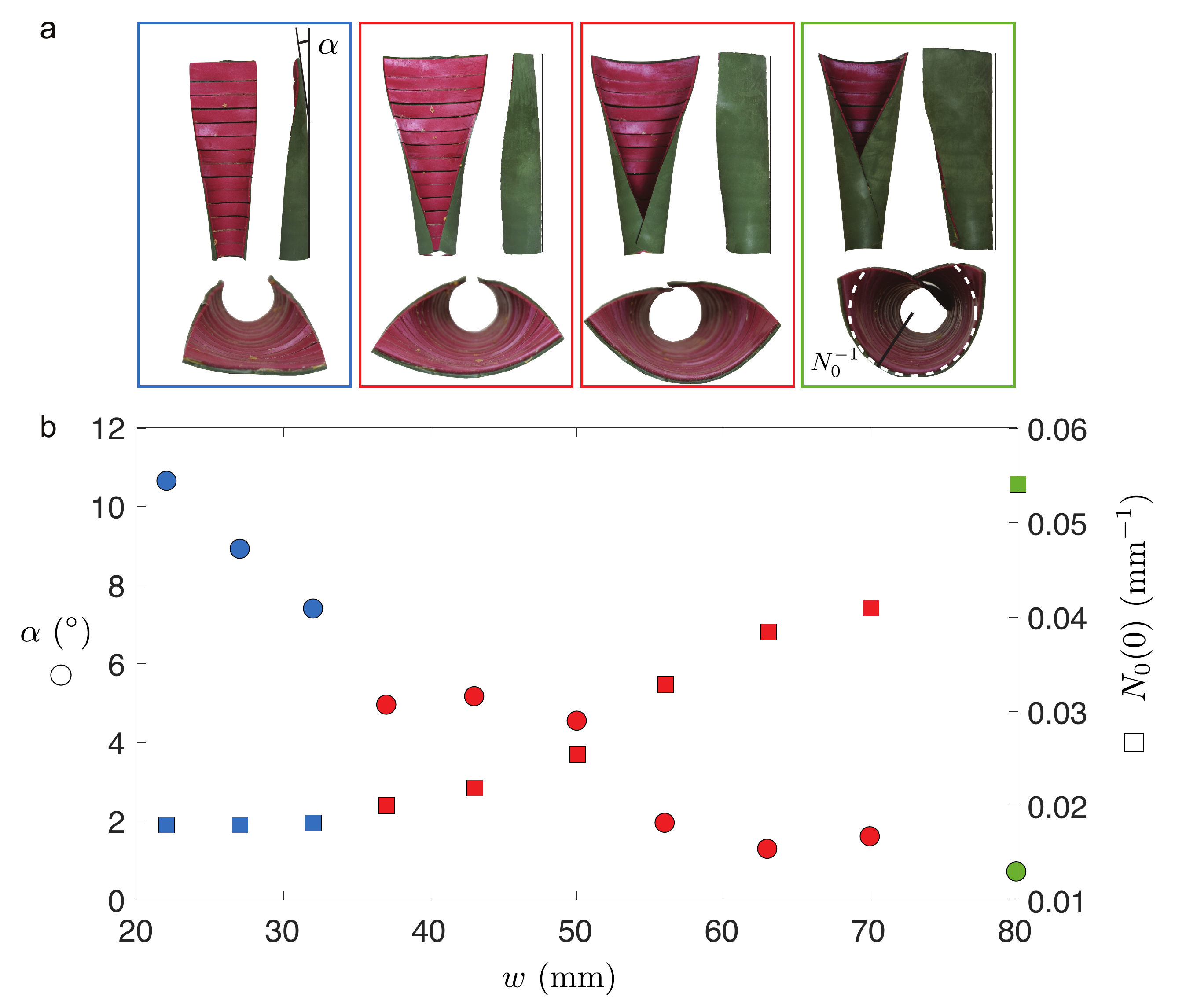}
    \caption{{\bf Shape transitions in ribbons with linearly varying transverse reference curvature}. (a) Evolution of the ribbon's morphology depending on the width. For narrow ribbons (left), the structures follow the  reference transverse curvature at stretching cost. The midline exhibits normal curvature, inducing a rotation angle $\alpha$. The resulting non-vanishing Gaussian curvature indicates stretching in the structure. For wide ribbons (right), the structure adopts a conical --- hence developable --- shape, with no curvature of the midline and with transverse curvature different from the reference. For intermediate widths (center, red boxes), ribbons present both Gaussian curvature and transverse curvatures different from the reference, indicating a mixed regime.
    (b) Midline rotation angle $\alpha$ and transverse curvature $N_0(0)$ as a function of the width. Colors indicate the regimes: when the end transverse curvature $N_0(0)$ follows its reference counterpart $\overline{N}_0(0)=0$, the ribbon is in a bending-dominated regime (blue). When the rotation angle of the midline $\alpha$, (and hence the Gaussian curvature) vanishes, the ribbon is in a stretching-dominated regime (green). Between these two asymptotic cases, a transition, colored in red, is observed. 
    Experimental parameters: $t=1.5$ mm, $\lambda=27$ mm and $\ell=90$ mm.
    }
    \label{figtrumpet}
\end{figure}
 \begin{figure}[h!]
    \centering
    \includegraphics[width=0.75\linewidth]{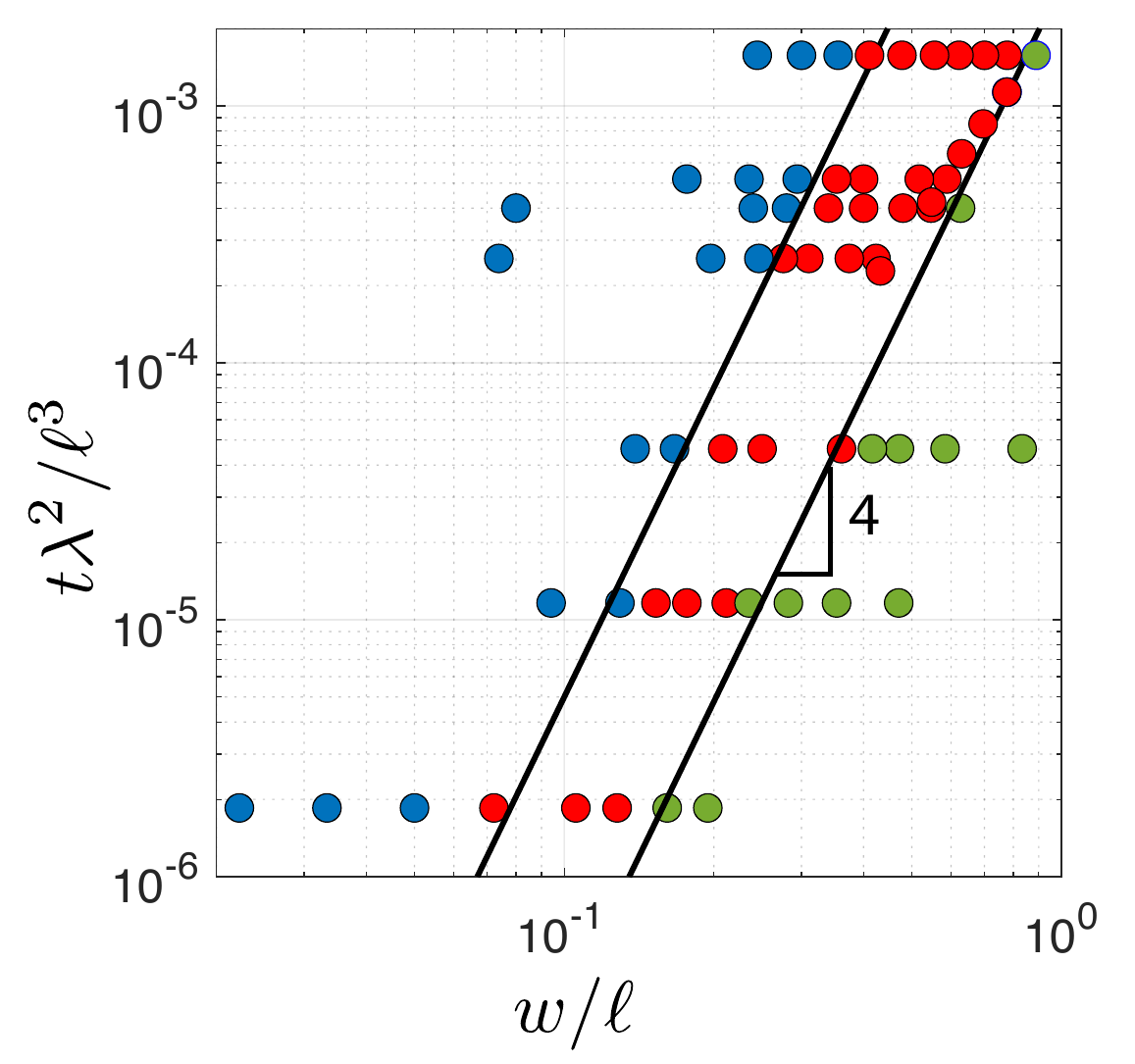}
    \caption{{\bf Phase diagram of ribbons with linearly varying transverse curvature}. Rescaled width $w/\ell$ as a function of the dimensionless quantity $t\lambda^2/\ell^3$. Colors indicate the regimes observed experimentally (the rationale for the coloring is presented in Fig.~\ref{figtrumpet}). The transition between regimes follows a power 4 law, as predicted by the theory (competition between $S_2$ and $B_0$ energy terms).
    In this regime, cones are observed in the thin limit (see Fig.~\ref{figtrumpet}), and we thus use the corresponding scaling of the energy.
    }
    \label{figtrumpetphase}
\end{figure}

The actual metric and curvature tensors postulated to find the improved scaling may be integrated to get the corresponding developable surface and compared to experiments. In Fig.\ \ref{figtrumpetiso}, different views of such a ribbon are presented and show very good agreement with the experimental ribbon, made by gluing prestretched rubber bands on double-sided tape. In this case, we chose $-\ell/2\le x_1 \le \ell/2$.

In cases where $\Nb_0(x_1)$ does not change its sign (i.e., when $x_1$ does not cross the origin), another candidate (observed experimentally) for the stretch-free configuration is a cone, for which the radius of curvature linearly varies ($N_0 = (\alpha + \beta x_1)^{-1}$ and $K= 0$).
In this case, MCP equations will provide corrections for $M_1$ and $L_2$ that will contribute to higher order of $B$.
Such cones have two degrees of freedom, the radius of curvature at the origin $\alpha$, and the rate at which it changes, $\beta$.
We can match $\alpha$ to the average of $\Nb$ (i.e. $\alpha \sim \lambda ^2 \ell^{-1}$) and $\beta$ to the average change rate  of $\Nb^{-1}$ (i.e $\beta \sim \lambda ^2 \ell^{-2}$).
Numerical minimization of the energy confirms this guess.
This results in $ B_{\text{cone}} \sim \dfrac{t^2\ell^2}{\lambda ^4}$
which is the same scaling as the naive bending energy estimation used in equation \eqref{eq:w4}. 
Such scaling appears to be asymptotically larger than $B_\delta$ (the bending content of the shape obtained from \eqref{eq:ab_perturb}).
However, taking into account the numerical prefactor, we get $\frac{B_{\text{cone}}}{B_\delta}\approx 0.07\frac{\ell}{w}$;
hence, whenever $w \gtrsim \ell/10$ we should expect a cone, which is indeed what we observe in our experiments. 
It should be emphasized that whenever the curvature changes sign, the cone's solution becomes completely unfavorable.

In Fig.~\ref{figtrumpet}, we show experimental realizations of such frustrated ribbons, with $0\leq x_1\leq \ell$, made by gluing prestretched (with linearly varying imposed strain) rubber strips on a passive rubber sheet (see \ref{ap:fab} for more details). Varying the width of the ribbon, we observe a shape transition, from a developable conical shape in the wide limit to a shape with Gaussian curvature, obeying locally the transverse curvature at stretching cost (Fig.\ \ref{figtrumpet}a). In order to quantify this shape transition, we measure both the transverse curvature at the origin $N_0(0)$ and the total rotation angle $\alpha$ of the midline $\m$. In the wide limit, the Gaussian curvature is expected to vanish everywhere and hence $\alpha=0$; such ribbons are then denoted in green. In the narrow limit, we expect the actual transverse curvature $N_0$ to follow the reference curvature, i.e, $N_0(0)=\Nb_0(0)=0$, and we denote such limit in blue. When none of these conditions are met, the ribbon is in an intermediate regime, denoted in red (Fig.~\ref{figtrumpet}b). 
Based on this rationale, we plot the phase diagram in Fig.~\ref{figtrumpetphase}, varying as much as possible the relevant geometrical parameters. The shape transition observed experimentally does obey the scaling law given by equation~\eqref{eq:w4}. The same scaling, as well as a non-vanishing rotation angle $\alpha$ of the midline, were reported in the case of the opening of a ribbon with an imposed curvature at one boundary~\cite{barois14}.

\subsection{Linearly varying longitudinal curvature: frustration between $S_0$ and $B_1$ and no transition}
\label{sec:invtrumpet}
In this section, we consider a ribbon with $-\frac{\ell}{2} \le x_1 \le \frac{\ell}{2}$ and the  following reference geometry (\figref{fig1}c):
\[
\ab = \smatrixII{1}{0}{1} 
\qquad
\bb = \smatrixII {\lambda^{-2} \, x_2}{0}{0}.
\]
The metric being flat, we have $\Kb = 0$. $\ab$ and $\bb$ also satisfy Gauss equation everywhere, since $\det\bb=0$.
The frustration thus stems from \MCP incompatibility. 
Imposing a stretch-free configuration, i.e., $\aac=\ab$, one may solve $\bac_0=\bb_0 = 0$, resulting in $\bac=0$, which corresponds to a flat ribbon.
Thus, assuming $\aac=\ab$ and $\bac_0=\bb_0$, we cannot have $\bac_1=\bb_1$, as $\Lb_1=\lambda^{-2} \ne 0$.
Since already $K_0=\Kb_0$ and $\bac_0=\bb_0$ are inconsistent with $L_1 = \overline{L}_1$, we have a competition between $S_0$ and $B_1$.
Therefore, the leading stretching term is of a lower order than the leading bending term (in the terminology of the previous section, $j=1>0=k$), hence \emph{stretching will always be the dominant term and we do not expect any phase transition}.
This result is confirmed by the experiments (see Fig.~\ref{fig1}e), where the ribbons considered remain identically flat for all parameters investigated, except at both ends of the ribbons, where boundary layers are observed.
Although there is no phase transition, the incompatibility implies that the energy is non-zero; indeed, since $L_1 \ne \Lb_1$, it scales like $t^2w^2$. This is a general lower bound for the elastic energy for MCP-incompatibility, as we discuss in the next section.

\subsection{Radial curvature: frustration between $S_1$ and $B_0$}
\label{sec:radial}

Consider a narrow section of an annulus in polar coordinates, with a geodesic curvature $\kgb$ and a constant reference curvature $\kappa$ in the radial direction (example $d$ in Fig.\ref{fig1}). 
Here, $x_1$, the long direction, is the (rescaled) annular coordinate, and $x_2$ is the radial coordinate (up to translation).
The reference geometry is then given by
\[
\ab = \smatrixII{(1-\kgb x_2)^2}{0}{1}
\qquad
\bb = \smatrixII{0}{0}{\kappa}.
\]
Again, the Gauss equation is satisfied by $\ab$ and $\bb$, but the \MCP equations are not.

\begin{figure}[h!]
    \centering
    \includegraphics[width=0.75\linewidth]{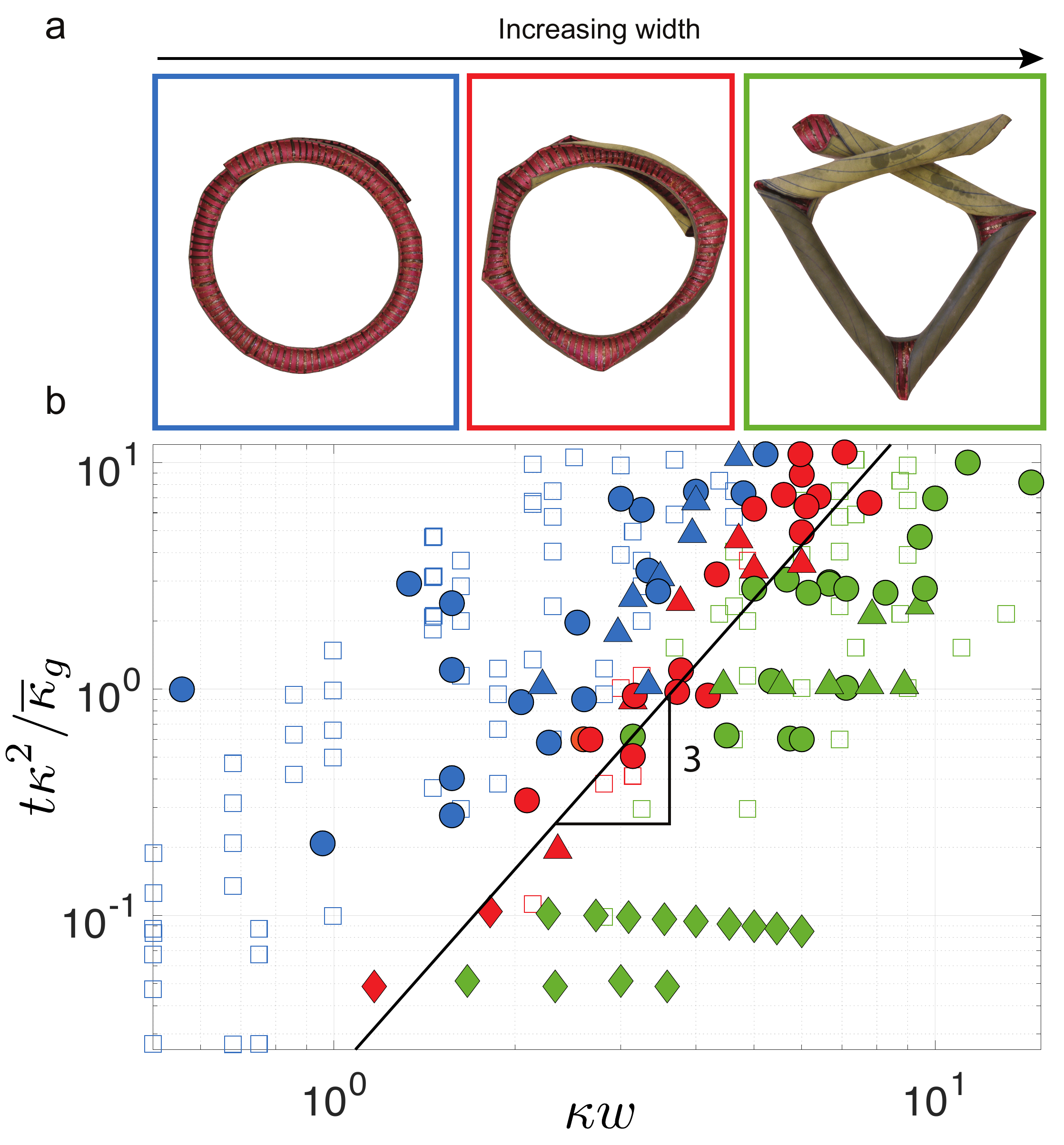}
    \caption{{\bf Shape transitions in curved ribbons with radial reference curvature}. (a) Different shape regimes observed experimentally: smooth toroidal shapes (in blue), which follow the reference radial curvature at a stretching cost; toroidal shapes with defects, through the buckling of the outer edge (in red); piecewise-tubular shapes, which obey the flat reference metric at a bending cost (in green). (b) Phase diagram, plotting $\kappa w$ as a function of  $t\kappa^2/\kgb$. The shape transition occurs along a slope 3, as predicted by the theory. The data are taken from Ref.~\cite{PRX}. Open squares correspond to numerical simulations, circles to latex bilayers, triangles to 3D-printed PLA ribbons and diamonds to latex-polypropylene bilayers.
    }
    \label{figtoriphase}
\end{figure}

We start by assuming $K=\Kb\equiv 0$ and trying to solve for $\bac$ with initial conditions $\bac_0 = \bb_0$ (i.e., $N_0=\kappa$ and $L_0=M_0=0$).
The zeroth-order Gauss equation is satisfied immediately.
The next order (which is $\Gn{1}$ and $\MCPn{0}$) is:
\[
\begin{split}
    \MCPn{0}&:\quad \left\{
    \begin{split}
        L_1 &= -\kgb \kappa\\
     M_1 &= 0 \\
    \end{split}
    \right.
    \\
\Gn{1}&:\quad \kappa L_1 = \Kb_1 = 0
\end{split}
\]
which is inconsistent.
Therefore, $\bac_0=\bb_0$ is incompatible with $K=\Kb$, and this inconsistency appears in $K_1$, since the zeroth-order Gauss equation is satisfied; thus, we have a frustration between $B_0$ and $S_1$.
If the configuration satisfies $\bac_0=\bb_0$, then $\MCPn{0}$ implies that $L_1=-\kgb \kappa$, and thus, from $\Gn{1}$ we have that $\Abs{K_1-\Kb_1}^2 = \Abs{\Nb_0 L_1}^2\sim \kgb^2\kappa^4$.
Therefore, in this case $S_1\sim \kgb^2 \kappa^4 w^6$.
Furthermore, $B_0\sim \kappa^2 t^2$ if the configuration satisfies $K_0 = K_1 = 0$ so we expect a transition $t\sim \kgb \kappa w^3$.
The same scaling was found in Ref.~\cite{PRX}, through a less systematic approach. It shows very good agreement with experimental data, as shown in Fig.~\ref{figtoriphase}.

As in Section~\ref{sec:trumpet}, we can improve the scaling by perturbing the reference geometry, and instead of considering $\bac_0=\bb_0$, we take $\bac_0 = \kappa \smatrixII{\delta^2}{\delta}{1}$ where $\delta$ is assumed to be small. 
Note that for every $\delta$, we maintain $K_0=0$ so $\Gn{0}$ is satisfied.
We search for a configuration with $\aac = \ab$ and whose curvature $\bac$ is $x_1$-independent.
$\bac$ must satisfy the GMCP equations identically, which under our assumptions read:
%we can find $\bac$ as  a solution to the GMCP equations with $\bac_0$ as the initial conditions:
\[
\begin{split}
    \text{MCP}&:\quad \left\{
    \begin{split}
        M'&=\dfrac{M \kgb}{1-x_2 \kgb} \\
 L'&=\dfrac{L \kgb}{x_2 \kgb-1}+N (\kgb (x_2 \kgb-1))
    \end{split}
    \right.
    \\
\text{G}&:\quad L N=M^2
\end{split}
\]
where the derivatives are taken with respect to $x_2$.
Substituting $N=M^2/L$ results with
\[
\left\{
\begin{split}
 M'&=\dfrac{M \kgb}{1-x_2 \kgb} \\
 \dfrac{(L^2)'}{2}&=\dfrac{L^2 \kgb}{x_2 \kgb-1}+M^2 (\kgb (x_2\kgb-1))\\
\end{split}
\right.
\]

\begin{figure}[h!]
    \centering
    \includegraphics[width=0.75\linewidth]{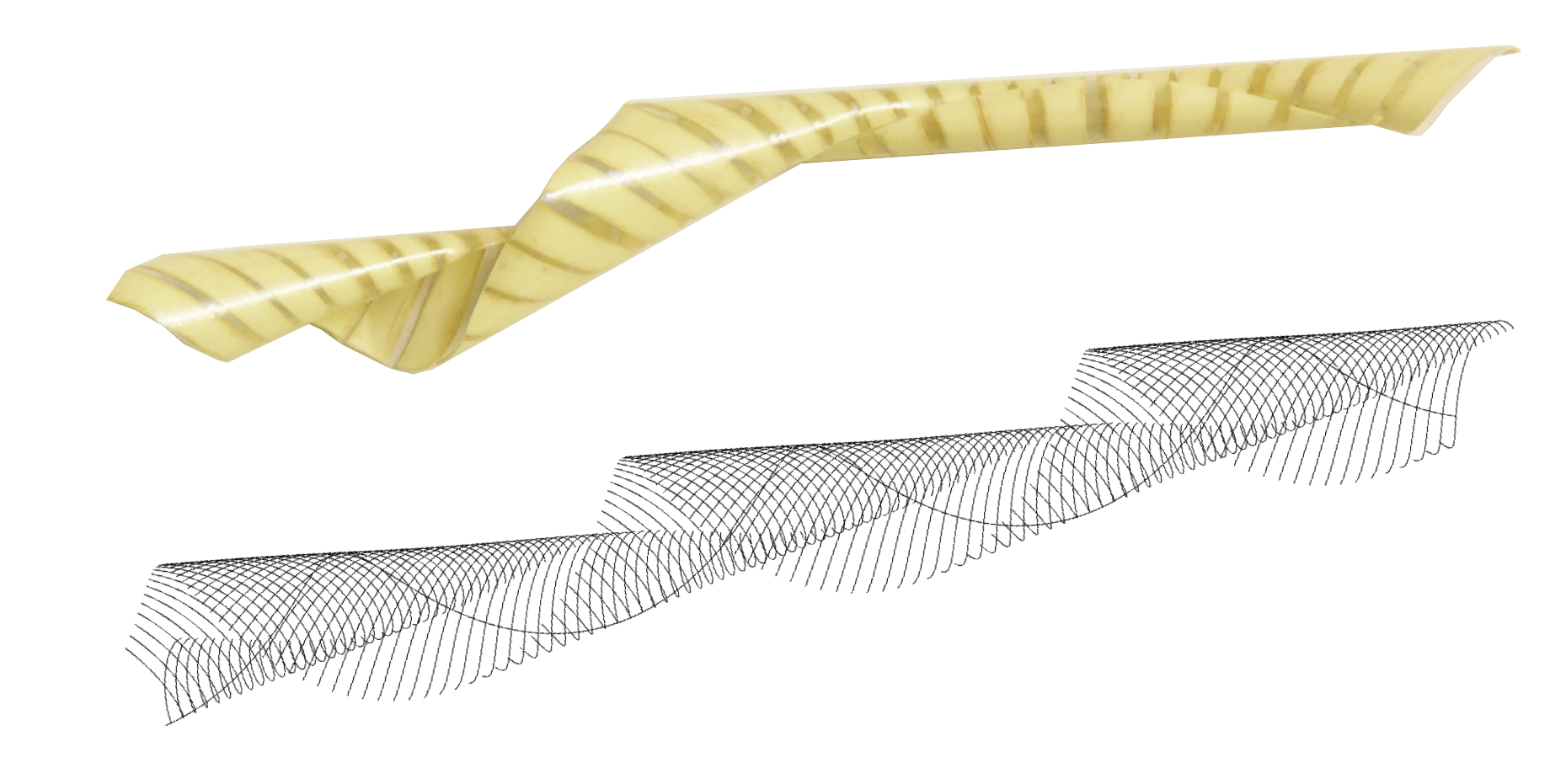}
    \caption{{\bf Helicoidal developable shape} Comparison between an experimental realization and the theoretical shape found by our perturbative approach. Helicoidal configurations are experimentally observed only when the total integrated geodesic curvature $\kgb\ell$ is sufficiently small, such that no defects appear between tubular portions.
    }
    \label{fighelicoidal}
\end{figure}

This system can be solved explicitly, yielding
%Solving these equations (which give an {\it exact} solution to the GMPC equations) yields
\[
\left\{
\begin{split}
 L^2&=\delta ^2 \kappa ^2((\delta ^2+1) (x_2 \kgb-1){}^2-1) \\
 M&=\dfrac{\delta  \kappa }{1-x_2 \kgb} \\
\end{split}
\right.
\]
and thus
\[
\bac = \delta\kappa\smatrixII
{\Brk{(\delta ^2+1)(x_2 \kgb-1)^2-1}^{\frac{1}{2}}}
{\brk{1-x_2 \kgb}^{-1}}
{\brk{1-x_2 \kgb}^{-1}\Brk{(\delta ^2+1)(x_2 \kgb-1)^2-1}^{-\frac{1}{2}}}.
\]
Note that since we {\it exactly} solved the GMPC equations, and our domain is simply-connected, $\aac=\ab$ and $\bac$ are valid metric and curvature tensors of a configuration that can be integrated from them.
By assuming $\delta \gg \kgb w$ we can expand $\bac \sim \kappa\smatrixII{\delta ^2}{\delta}{1-\kgb w \delta^{-2}}$, whose corresponding bending energy is minimized for $\delta \sim (w \kgb)^{1/3}$.
Therefore, this ansatz results in a bending energy of $B_\delta\sim t^2 (\kappa ^2 (w \kgb)){}^{2/3}$ and hence a transition of $t\sim \kappa  w^{8/3} \kgb^{2/3}$ (the competing stretching term is still $S_1 \sim w^6 \kappa^4\kgb^2$). 
This provides a tighter bound, changing the exponent of $w$ in the transition by $\gamma = 1/3$ ($t\propto w^{8/3}$ instead of $t\propto w^3$).
Integrating the shape given by this actual geometry, we get a helicoidal configuration with a conical aspect (due to the geodesic curvature of the midline). 
In experiments, we mostly observe piecewise-tubular ribbons, in the stretching-dominated (wide) regime (see Fig.~\ref{figtoriphase}a), rather than this helicoidal configuration. 
While the piecewise-tubular shapes do not change the naive scaling $t\sim w^3$ (since the corners in-between two tubular sections need to fully unbend over a finite area, see \cite{PRX}), they strongly reduce the prefactor of the bending energy, thus they may be preferable over the helicoidal shape, at least in some regimes.
Nevertheless, in the case of relatively small portions of ribbons, where no corners appear, we do observe this helicoidal configuration near one edge (see Fig.~\ref{fighelicoidal}), showing thus, the relevance of our analysis.
In any case, transition between bending-dominated and stretching-dominated approximately at $t\sim w^3$ is indeed in good accordance with experiments and numerical simulations, as shown in Fig.~\ref{figtoriphase}b.

\subsection{Azimuthal curvature: frustration between $S_g$ and $B_1$ and no transition}
\label{sec:azimuthal}

Consider the same section of an annulus in polar coordinates, as in Section~\ref{sec:radial}, but now with 
a constant reference curvature $\kappa$ in the {\it azimuthal} direction (\figref{fig1}e).
The reference geometry is then given by
\[
\ab = \smatrixII{(1-\kgb x_2)^2}{0}{1}
\qquad
\bb = \smatrixII{\kappa (1-\kgb x_2)^2}{0}{0},
\]
and again, the incompatibility only stems from the \MCP equations.

\begin{figure}[ht!]
    \centering
    \includegraphics[width=0.8\linewidth]{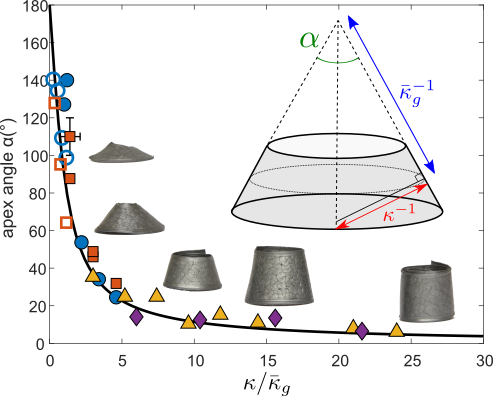}
    \caption{{\bf Shape of a ribbon with geodesic and azimuthal curvatures}. 
        The apex angles for different conical shapes of azimuthally curved ribbons of varying width (5, 7, 10, 20 mm, corresponding to blue circles, red squares, yellow triangles, and purple diamonds, respectively) and thickness (0.3mm and 0.6mm, filled and empty markers, respectively).
        The solid line is the graph of the relation $\alpha = \arctan(\kgb / \kappa)$ in a perfect cone, as shown in the inset. Here $\kappa$ is the normal curvature along the midline. 
        All cones collapse to this trend indicating no shape transition.
        Along the line, several examples of 4D printed samples are shown.
    }
    \label{fig:az_curv}
\end{figure}

We start in the stretch-free limit, hence $K=\Kb=0$.
Fixing $\bac_0 = \bb_0$ invokes no frustration in $\Gn{0}$.
In the next order, $\MCPn{0}$ and $\Gn{1}$ yields
\[
\begin{split}
    \MCPn{0}&:\quad \left\{
    \begin{split}
        L_1 &= -\kgb \kappa\\
     M_1 &= 0 \\
    \end{split}
    \right.
    \\
\Gn{1}&:\quad N_1 = 0 
\end{split}
\]
which is consistent.
However we immediately see that we cannot satisfy $\bac_1 = \bb_1$ since $L_1=-\kgb \kappa$, which is forced by the previous terms through $\MCPn{0}$, is different from $\Lb_1=-2\kappa\kgb$.
This leads to a competition between the geodesic curvature (and hence $S_g$) and $B_1$.
Since the energetic cost of $S_g$ is of order $w^2$ and $B_1$ is of order $t^2w^2$, the geodesic curvature is indeed satisfied, %and which asymptotically favours $S_g$ and 
the resulting actual geometry is given by:
\[
\aac = \smatrixII{(1-\kgb x_2)^2}{0}{1}
\qquad
\bac = \smatrixII{\kappa  (1-x_2 \kgb)}{0}{0},
\]
which indeed satisfy the GMCP equations identically. 
Note that this geometry is optimal even if we only assume $\kg = \kgb$ and $\bac_0 = \bb_0$, hence it is preferable even in the bending-dominant regime.
Therefore in this case there is no shape transition and the configuration has an energy of order $t^2w^2$.
Such configuration corresponds to a cone with an apex angle of $\arctan(\kgb / \kappa)$ (see inset of \figref{fig:az_curv}).
Experiments done using 4D printing supports these results (\figref{fig:az_curv}).

\section{Discussion and conclusion}

\begin{table}[ht!]
\centering
\begin{tabular}{|c|l l|c|c|} 
\hline
& \multicolumn{2}{c|}{Reference geometry} & Competing terms & Transition scaling \\ 
\hline
& & & &\\
a & $\ab = \smatrixII{1}{0}{1}$ & $\bb = \smatrixII {0}{\kappa}{0}$ 
& $S_0$ and $B_0$ & $t\sim \kappa w^2$\\
& & & &\\
b & $\ab = \smatrixII{1}{0}{1} 
$&$
\bb = \smatrixII {0}{0}{\lambda^{-2} \, x_1}$ & $S_2$ and $B_0$ & $t\sim \ell^{-1}\lambda^{-2}w^4$\\
& & & &\\
c & $\ab = \smatrixII{1}{0}{1} 
$&$
\bb = \smatrixII {\lambda^{-2} \, x_2}{0}{0}$ & $S_0$ and $B_2$ & No transition\\
& & & &\\
d & $\ab = \smatrixII{(1-\kgb x_2)^2}{0}{1} 
$&$
\bb = \smatrixII {0}{0}{\kappa}$ & $S_1$ and $B_0$ & $t\sim \kgb\kappa w^3$\\
& & & &\\
e & $\ab = \smatrixII{(1-\kgb x_2)^2}{0}{1} 
$&$
\bb = \smatrixII {\kappa(1-\kgb x_2)^2}{0}{0}$ & $S_g$ and $B_1$ & No transition\\
& & & &\\
 \hline
\end{tabular}
\caption{{\bf Summary of the scalings obtained in Section~\ref{sec:examples}, using the asymptotic theory presented in Section~\ref{sec:asymptotic}}. 
The lettering of the examples corresponds to the one introduced in Fig.~\ref{fig1}.
Recall that $S_g, S_k$ (for $k \ge 0$), and $B_j$, are the result of deviations in the geodesic curvature, the Gaussian curvature (of order $k$) and the curvature tensor (of order $j$), respectively, see \eqref{eq:energy_1D}.
}
\label{tab}
\end{table}
We have presented a general asymptotic theory for frustrated elastic ribbons with arbitrary geometry. 
Given an incompatible reference geometry, we expand the curvature and metric tensors with respect to the width $w$ of the ribbon, identified as a small parameter.
We solve order by order the compatibility equations until we encounter inconsistency, allowing us to find the dominant competing terms in bending and stretching. 
From them we deduce whether a shape transition is expected, and when it is the case, bounds on the scaling of the transition are found.
We successfully apply this systematic approach to five different incompatible ribbons, resulting in competition between different terms and hence different scaling-laws.
The main results of our paper are summarized in Table~\ref{tab}.

For specific cases in which the reference curvature along the midline $\bb_0$ is inconsistent with the reference metric, the scaling may be refined by perturbing $\bb_0$ in order to find zero-stretching configuration with bending energy $\ll t^2$.   
In fact, the results of \cite{FHMP16b} imply that this is always possible whenever $\Kb=0$ and there is no zeroth-order Gauss incompatibility (i.e., when $\det \bb_0 = 0$).
%estimating more accurately the energy involved in the asymptotic configuration.
Nevertheless, such refinement is limited to reducing the power of $w$ in the scaling of the transition in less than 1.

The bending-dominated limit corresponds to a regime where the ribbon behaves like a rod, with significant stretching (though the midline itself, of course, remains unstretched), whereas the stretching-dominated limit may be seen as a plate-like, stretch-free regime, which corresponds to most studies in the literature.

In this work, we assumed isotropy of the material response for the sake of simplicity. 
However, while anisotropy would certainly affect the precise shape selection, we argue that the scaling found in this article should not be impacted.
Moreover, in cases where the constitutive relation is not Hookean, a similar approach could be used, but the power-laws would be modified accordingly.

There are many more cases of incompatibility, beyond the ones considered here, that may be studied within this framework.
In particular, higher-order Gauss frustration (i.e., Gauss incompatibility that would not appear on the ribbons' midline), could result in a competition between $S_1$ and $B_1$ (for first-order Gauss incompatibility) and would therefore lead to the same scaling as in the case of zeroth-order Gauss incompatibility (Example a in Table~\ref{tab}). 
In such case, we anticipate the shape transition to be more subtle as it involves higher terms in the expansion of the geometry.
These high-order Gauss incompatibilities were recently numerically investigated in \cite{huang2018}. 
Our approach may help to analytically capture the rich morphologies and shape transitions (from twist to edge waving) observed.
Other cases may still arise in pure MCP-incompatibility --- for example, the reference curvatures in examples b and c in Table~\ref{tab} can be viewed as 90 degrees rotations of each other; for other rotations one expect to have a competition between $S_1$ and $B_1$, resulting in $t\sim w^2$, hence this scaling can appear when $\ab$ and $\bb$ are Gauss-compatible.
Yet more cases can occur due to combinations of Gauss and MCP-incompatibilities, or when $\ab$ is non-flat.

Regardless of whether and where a transition occurs, whenever there is zeroth-order MCP-incompatibility, the bending energy will be at least of order $t^2w^2$ in all regimes: indeed, the incompatibility in the $\textup{MCP}_0$ equation implies that if $\bac_0 = \bb_0$, then $\bac_1$ cannot be $\bb_1$, resulting in the bending term $B_1$, which scales like $t^2 w^2$, being non-zero.

Finally, we note that a key aspect of our analysis is treating the \GMCP compatibility equations as an \textit{initial value problem} for $\bac$: the metric $\aac=\ab$ is given and determines the equations for $\bac$, and the curvature along the midline $\bac_0 = \bb_0$ (which satisfies $\det \bb_0 = 0$) is the initial condition.
The solvability of this system depends on the initial data $\bb_0$, and when it is not solvable, as in examples b and d, there are phase transitions with non-standard scalings (whereas in examples c and e we solve this system explicitly).
As far as we know, 
%this is the first time when the compatibility equations are treated as an initial value problem, and 
exact solvability conditions for the initial data $\bb_0$, in the case of $\Kb=0$ considered here, have not been derived (beyond the necessary Gauss compatibility $\det \bb_0=0$).

\section*{Acknowledgments}
This research was supported by the USA--Israel binational science foundation, Grant No. 2014310.
I. L. is grateful to the Azrieli Foundation for the award of an Azrieli Fellowship.
E. S. acknowledges support from the Lady Davis Fellowship Trust.
C. M. was partially supported by ISF-grant 1269/19

\bibliography{references_vr2.bib}

\appendix
\section{Justification of the semi-geodesic expansion of $\aac$}
\label{ap:metric}
We start by expanding $\aac\equiv \smatrixII{e}{f}{g}$ around the midline:
\[
\aac = \smatrixII
{e_0 + e_1 x_2 + \dfrac{1}{2} e_2 x_2^2+...}
{f_1 x_2 + \dfrac{1}{2} f_2 x_2^2+...}
{1 + g_1 x_2 + \dfrac{1}{2} g_2 x_2^2+...}
\]
We start by noting that $f_1,g_1 \ne 0$ will have energetically cost of order $w^2$ (the same argument for $\kg=\kgb$).
To estimate the effect of higher order terms on the compatibility, we look at their contribution to the Gaussian curvature.
The Gaussian curvature associated with $\aac$ is given by the Brioschi formula:
\[
K=\dfrac{\left|\begin{array}{ccc}
-\dfrac{1}{2} e_{v v}+f_{u v}-\dfrac{1}{2} g_{u u} & \dfrac{1}{2} e_{u} & f_{u}-\dfrac{1}{2} e_{v} \\
f_{v}-\dfrac{1}{2} g_{u} & e & f \\
\dfrac{1}{2} g_{v} & f & g
\end{array}\right|-\left|\begin{array}{ccc}
0 & \dfrac{1}{2} e_{v} & \dfrac{1}{2} g_{u} \\
\dfrac{1}{2} e_{v} & e & f \\
\dfrac{1}{2} g_{u} & f & g
\end{array}\right|}{\left(e g-f^{2}\right)^{2}}.
\]
It follows that expanding $\aac$ order-by-order, $e_k$ contribute already to $K_{k-2}$ while $f_k$ and $g_k$ contributes only to $K_{k-1}$ and higher terms.
Nevertheless, they all contribute to $S_{k-2}\propto w^{2k}$ and $e$ is enough to satisfy the Gauss equation.
As a consequence, while $e_k$ is involved in the $(k-2)$-order Gauss equation, $\Gn{{k-2}}$, $f_k$ and $g_k$ are involved only in $\Gn{{k-1}}$ and higher. 
Since $\Gn{{k-1}}$ involves $\bac_i$ for $i$ up to order $k-1$, it follows that $f_k$ and $g_k$ are competing with $\bac_{k-1}$, whose energetic contribution is $B_{k-1} \propto t^2 w^{2(k-1)} \ll S_{k-2}$; thus it is favorable to set $f_k=g_k=0$ (which minimizes $S_{k-2}$) and pay some bending.
Similarly, $\MCPn{k}$ (the MCP equations of order $k$) are constraining $\bac_k$ and $\aac_{k-1}$, but the contribution of $\aac_{k-1}$ to the energy is always higher than $\bac_k$.
Thus, for each order $\aac_k$ we find $f_k=g_k=0$ to be energetically favorable, and the possible competition is between choosing $e$ that minimizes the stretching and $\bac$ that minimizes the bending.

\section{Experimental methods}
\label{ap:fab}
\begin{figure}[h!]
    \centering
    \includegraphics[width=\linewidth]{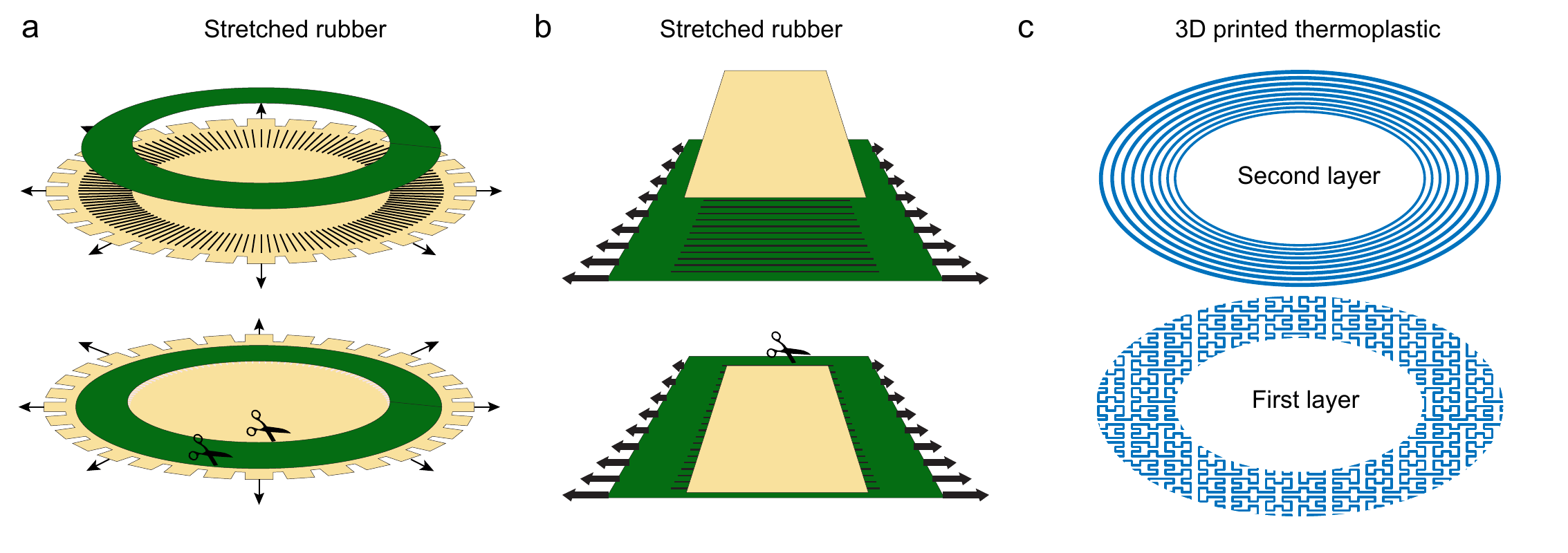}
    \caption{{\bf Fabrication of frustrated ribbons}. 
    (a) Fabrication of curved ribbons with radial constant reference curvature: a rubber sheet is radially cut and stretched and a passive sheet is glued on it.
    A curved ribbon is then cut from the structure.
    (b) Fabrication of straight ribbons with linearly varying curvature: a set of parallel lines are cut from a rubber sheet, defining strips.
    The sheet is then stretched along the strips, the imposed strain varying linearly along the direction perpendicular to the strips.
    A passive sheet is glued on the stretched structure and a ribbon of desired dimension is then cut from the bilayer structure, the linearly varying pre-stretch inducing a linearly varying reference curvature.
    (c) Fabrication of curved ribbons with azimuthal constant reference curvature. In this case, stretched rubber sheets is impractical since the direction of curvature varies.
    We thus use 3D-printed thermoplastic (PLA) ribbons: a layer made of concentric circles is printed on top of a  first layer filled with a Hilbert curve.
    When heated, the first layer is mostly passive whereas the second one contracts along the azimuthal printing direction~\cite{an18}, inducing the sought reference curvature.
    }
    \label{figfab}
\end{figure}

We fabricate the geometrically frustrated ribbons using two distinct strategies, depending on the configurations. 
When the directions of non-vanishing reference curvature are straight line as in the case of Fig.~\ref{fig1}a--d, we prestretch fine strips --- made by laser-cutting rubber sheets --- along those directions (Fig.~\ref{figfab}a--b).
A second passive rubber sheet is then glued on the first prestretched sheet. When the glue is dry, ribbons of desired dimensions are cut from the bilayer rubber sheet.

When the directions of non-vanishing reference curvature are curved however, this method is not suitable. We thus turn to 3D printed structures using PLA. When printed sufficiently fast and thin, this thermoplastic material solidifies in a state where the polymeric chains are straightened and aligned because of the extensional flow through the nozzle~\cite{van2017programming}. When heated above the glass transition temperature, the material thus contracts along the printing direction. In order to program azimuthal reference curvature we print bilayer structures: the first layer is relatively thick and filled with an isotropic pattern (Hilbert curve) and is thus mostly passive. The second layer is made of concentric circles (Fig.~\ref{figfab}c) and contracts azimuthally when heated, inducing an azimuthal reference curvature to the ribbon.

\end{document}